\documentclass[a4paper,11pt]{article}
\pdfoutput=1 

\usepackage{jheppub} 

\usepackage[T1]{fontenc}

\title{\boldmath Complexity, Entropy, and Markov Chains}

\author{Zhou Shangnan}
\affiliation{Stanford Institute for Theoretical Physics and Department of Physics, Stanford University, \\ Stanford, CA 94305, USA}

\emailAdd{snzhou@stanford.edu}

\abstract{We develop a theory of classical complexity. We study the relations between classical complexity and entropy, and conjecture that in an isolated system, classical absolute complexity always tends to grow, until it reaches its maximum. We calculate some exact closed-form expressions of the growth of average classical complexity over time in some concrete models, and gain further insights of both classical and quantum complexity by using the theory of Markov chains. }

\begin{document} 
\maketitle
\flushbottom

\section{Introduction}
\label{sec:intro}

Complexity theory is an interesting but difficult subject with growing attention \cite{brown_second_2018, susskind_switchbacks_2014,brown_quantum_2017,lin_cayley_2019,susskind_three_2018}. It is interesting because it is connected to many fascinating questions in holography and black hole physics. For example, it is conjectured that quantum computational complexity of a holographic state is given by the classical action of a region in the bulk \cite{brown_complexity_2016,brown_complexity_2016-1, susskind_computational_2014,stanford_complexity_2014,brown_case_2018,goto_holographic_2018}. It is difficult because there is a lack of mathematical tools to analyze it, so that it is hard to give concrete arguments and proofs. 

The main focus of this paper is to develop a concrete theory of classical complexity, with the hope that some intuitions, techniques, and results can be transferred to the theory of quantum complexity, since classical complexity is essentially quantum complexity without entanglement. The main part of the paper can also be considered as providing arguments for the existence of a thermodynamic theory of classical complexity, which is in complementary to \cite{brown_second_2018}. 

We begin by defining classical complexity and classical absolute complexity, as well as the so-called "bit model", "trit model", and "n-dit" model (section \ref{sec:2}). We then study the relations between classical complexity and entropy. We show that those states with maximal classical absolute complexity are also the states with maximal entropy, and vice versa. We then conjecture the second law of classical absolute complexity. In an isolated system, classical absolute complexity always tends to grow, until it reaches its maximum (section \ref{sec:3}). A concrete example is the Bit model, in which larger absolute complexity corresponds to larger entropy. We calculate some exact closed-form expressions of the growth of average complexity over time in the bit and trit models, and also give a good analytical estimate in the n-dit models. Moreover, we adopt mathematical tools of Markov chains \cite{grinstead_grinstead_2009, walsh_knowing_2012} to study the evolution of our systems, which makes it possible to calculate the average time it takes to reach the maximal classical absolute complexity, i.e. the first passage time to a state in the complexity equilibrium. It is also straightforward to justify that the recurrence time is exponential in $K$, the size of the system, from a Markovian perspective (section \ref{sec:4}). We suggest that these tools can also bring insights to the theory of quantum complexity (section \ref{sec:5}). Furthermore, we discover that the Bit model is dual to the Ising model \cite{kardar_statistical_2007} at infinite temperature, which hints towards a first law of classical complexity (section \ref{sec:6}).

\section{Definitions of classical complexity}
\label{sec:2}

While quantum complexity is about qubits, naturally classical complexity is about bits. As we all know, a bit can take values 0 and 1. For a single bit, a reversible operation we can do is

\begin{equation}
    T_+(0) = 1, \ T_+(1) = 0 
\end{equation}

Its inverse, which can be denoted as $T_-$, is actually identical to $T_+$. This operation is essentially a bit flip operation. 

A generalized version of bit is dit. An n-dit can take values 0, 1, ... , $n - 1$.  For a single n-dit, a reversible operation we can do is

\begin{equation}
\label{eq:2:2}
    T_+(i) = i + 1, \  i \neq n - 1; \  T_+(n - 1) = 0 
\end{equation}

Its inverse $T_-$ is

\begin{equation}
\label{eq:2:3}
    T_-(j) = j - 1, \  j \neq 0; \  T_-(0) = n - 1
\end{equation}

$T_+$, $T_-$, along with the identity operation $I$, form a group, denoted by $G$. $X = \{ 0, 1, ..., n - 1 \}$ is a set. Equations \eqref{eq:2:2} and \eqref{eq:2:3} define the group action on $X$.

Starting from an n-dit with value 0, we can apply $T_+$ or $T_-$, or a combination of the two, to reach any possible value of the n-dit. The individual state of an n-dit is defined by its value.

Now suppose we have $K$ n-dits. The system can be represented by a string of length $K$, which is a list of the values of the n-dits. The configuration state of the system is defined by this string: $\Phi_x = ( x_1, \ x_2, \ ... \ x_K )$, where each $x_i$ represents the value of the $i$th n-dit. We use the word state to refer configuration state when no confusion is caused.

We assume $n | K$, since this doesn't affect the big picture. A general operation $O$ on these n-dits changes the values of the n-dits. Since the n-dits are independent of each other (unlike the quantum case when the qubits can be entangled), $O$ is a tensor product of $T_+$, $T_-$ and $I$, and we simplify our discussion by restricting to the case that we only operate on one n-dit at a time. The growth of classical complexity, which will be studied later in section \ref{sec:4}, will not be fundamentally different if we get rid of this restriction. This can be thought of as an analog of quantum complexity, where as long as we have a universal gate set, it doesn't matter which specific gate set we choose. Here, since our operators make it possible to reach all possible configuration states, we essentially have a "universal gate set".

For reference purposes, when we operate on bits, we call our model "the bit model"; when we operate on trits, i.e. 3-dits, we call our model "the trit model"; when we operate on general n-dits, we call our model "the n-dit model".

We first define the classical relative complexity between state $A$ and state $B$ as the minimum operations needed to go from $A$ to $B$.

Now we need to find some good "reference point" to define classical complexity, i.e., what kind of states are the simplest? Well, by intuition, those states whose all n-dits have the same value are the simplest. For example, $(0,0,0,0,...,0)$ and $(1,1,1,1,...,1)$. For reference purposes, we call a state whose all n-dits have the same value $i$ as the "all-i" state.

One difficulty we then encounter is that we have multiple different "simplest states". By symmetry, these "simplest states" are all on the equal footing, and we either favor one of them to be the unique "simplest state" while standing the consequences of the breaking symmetry, or we find a way to accommodate multiple different "simplest states" while making sure complexity is well-defined.


This suggests two different definitions of classical complexity:

\begin{flushleft}
\textbf{Definition of Classical Complexity} We pick an "all-i" state as the simplest state. We define classical complexity as the minimum operations needed to go from the simplest state to the desired state, denoted by $C$.
\end{flushleft}

In this paper, since there is no fundamental difference between the "all-zero" state and other "all-i" states, we usually pick the "all-zero" state as the simplest state.

\begin{flushleft}
\textbf{Definition of Classical Absolute Complexity} All the "all-i" states are considered the simplest state. We define classical absolute complexity as the minimum operations needed to go from a simplest state to the desired state, denoted by $C_{abs}$.
\end{flushleft}

Classical complexity is essentially the relative complexity between the given simplest state and the desired state, which in some sense is a more "local" definition. Classical absolute complexity is more "global" since we need to compare our desired state with every simplest state.



In our case with $K$ bits, suppose the given simplest state is the "all-zero" state, then we have a well-defined complexity and absolute complexity. Their relation is 

\begin{equation}
    C_{abs} = \min{(C, K-C)} 
\end{equation}

If at each time step, we only operate on one bit, 

\begin{equation}
    \max(C_{abs}) = \frac{1}{2} \max(C) = \frac{K}{2}
\end{equation}
where $C$ equals to the number of "1"s we have in the configuration state.

For a single n-dit, if we define the state with value 0 as the given simplest state, i.e., with complexity zero. 
Then for an n-dit with value $j$, the corresponding classical complexity is 

\begin{equation}
    C(j) = \min (j, n-j)
\end{equation}

This implies that 

\begin{equation}
    C(j) = C(n-j)
\end{equation}

For a system with $K$ n-dits, if at each time step, we only operate on one n-dit, then the complexity of the system is the sum of the complexity of each n-dit. 


\section{Complexity, absolute complexity, and entropy}
\label{sec:3}

\subsection{States with maximal entropy = states with maximal absolute complexity}

For our system with $K$ n-dits ($n|K$), suppose there are $a_i$ n-dits with value $i$, we would like to know which distribution of values is the most probable. 

Since every n-dit is assigned to a value,

\begin{equation}
    \sum_{i=0}^{n-1} a_i = K
\end{equation}

We denote the number of configuration states for a given set of $a_i$'s by $N(a_0, a_1, ... , a_{n-1})$. 

\begin{equation}
    N(a_0, a_1, ... , a_{n-1}) = \frac{K!}{\prod_{i = 0}^{n -1} (a_i!)}
\end{equation}

The corresponding entropy is 

\begin{equation}
\label{entropy}
    S(a_0, a_1, ... , a_{n-1}) = \ln N(a_0, a_1, ... , a_{n-1})
\end{equation}

Eq.\eqref{entropy} is maximized when $a_0 = a_1 = a_2 = ... = a_{n-1} = \frac{K}{n} $, which can be proved by induction. An intuitive understanding is that $\prod_{i = 0}^{n -1} (a_i!)$ is a product of $K$ integers, and changing the values of $a_i$'s from $\frac{K}{n}$ to other values is equivalent to replacing some smaller integers in the product by larger integers.

Hence, the most probable value distribution is that $a_0 = a_1 = a_2 = ... = a_{n-1} = \frac{K}{n} $, with 

\begin{equation}
   N_{max} = \frac{K!}{(\frac{K}{n}!)^n} \approx n^K 
\end{equation}

\begin{equation}
    S_{max} \approx K \ln n
\end{equation}
for large $K$.

For an arbitrary $n$, since we define the simplest states as those states whose n-dits all have the exactly same value, each system has $n$ different simplest states. For a configuration state satisfying $a_i = \frac{K}{n}$ for any $i$, we calculate its corresponding complexity and absolute complexity.

If $n$ is odd,

\begin{equation}
    C = C_{abs} = \frac{K}{n} \sum_{i = 1}^{\frac{n - 1}{2}} 2i = \frac{n^2 -1}{4n} K
\end{equation}

If $n$ is even,

\begin{equation}
    C = C_{abs} = \frac{K}{n}\Big(\sum_{i=1}^{\frac{n}{2}-1} 2i + \frac{n}{2} \Big) = \frac{n}{4} K
\end{equation}

These configuration states actually have the maximal possible absolute complexity. This is equivalent to: for any configuration state $\Phi$ with $a_i$ n-dits taking value $i$, $C_{abs} (\Phi) \leq \frac{n^2 -1}{4n} K$ if $n$ is odd, $C_{abs} (\Phi) \leq \frac{n}{4} K$ if $n$ is even, and the equality holds only when $a_i = \frac{K}{n}$ for any $i$.

\textit{Proof}. There are $n$ different simplest states: the state whose all n-dits have value 0, the state whose all n-dits have value 1, ... , the states whose all n-dits have value $n -1$. We denote the complexity of $\Phi$ calculated with respect to the state whose all n-dits have value $j$ by $C_j (\Phi)$. $f (\Phi) = \sum_{j = 0}^{n - 1} C_j(\Phi)$ is a constant function since the summation puts all values on an equal footing. Since $C_{abs} (\Phi) = \min \big(C_0(\Phi), C_1(\Phi), ... , C_{n-1}(\Phi) \big)$, we get the maximal $C_{abs} (\Phi)$ when $C_0(\Phi) = C_1(\Phi) = ... = C_{n-1} (\Phi) = \frac{1}{n} f(\Phi)$. This holds only when $a_i = \frac{K}{n}$ for any $i$.

 A random configuration state evolves towards such states with maximal absolute complexity, since these states correspond to the largest entropy. We call these states at equilibrium.

\subsection{Complexity, entropy, and size of the system}

Suppose for our system with $K$ n-dits ($n|K$), there are $a_i$ n-dits with value $i$, and each $a_i$ is large, then by Stirling's approximation
\begin{equation}
    S(a_0, a_1, ... , a_{n-1}) = -\sum_{i=0}^{n-1} a_i \ln \frac{a_i}{K} 
\end{equation}

Let $q_i = \frac{a_i}{K}$, then $\sum_{i=0}^{n-1} q_i = 1$, and we can rewrite the entropy as a function of $q_0$, $q_1$, ... , $q_{n-2}$ and $K$. In the following equations, we implicitly use $q_{n-1} = 1 - \sum_{i=0}^{n-2} q_i$ for simplicity purpose.

\begin{equation}
    S(q_0, q_1, ... , q_{n-2}, K) = - \sum_{i=0}^{n-1} (q_i \ln q_i) K
\end{equation}

If the given simplest state is the "all-zero" state, then the corresponding classical complexity is

\begin{equation}
    C(a_0, a_1, ... , a_{n-1}) = \sum_{i=0}^{n-1} \min(i, n-i) \ a_i 
\end{equation}
which is equivalent to 

\begin{equation}
    C(q_0, q_1, ... , q_{n-2}, K) =  \sum_{i=0}^{n-1} \min(i, n-i) \ q_i K
\end{equation}

If we double the size of the system with the value distribution fixed, i.e. double K while fixing the $q_i$'s, then both entropy and complexity will be doubled. Of course, from the definitions, we expect these quantities to be extensive.




\subsection{Second law of classical absolute complexity}

A special case is that when $n = 2$, $a_1 = C$, $a_0 = K - C$, 

\begin{equation}
     S(a_0, a_1) = S(C, K) = \ln \binom{K}{C} = \ln \binom{K}{C_{abs}}
\end{equation}

\begin{figure}[tbp]
\centering
\includegraphics[scale = 0.75]{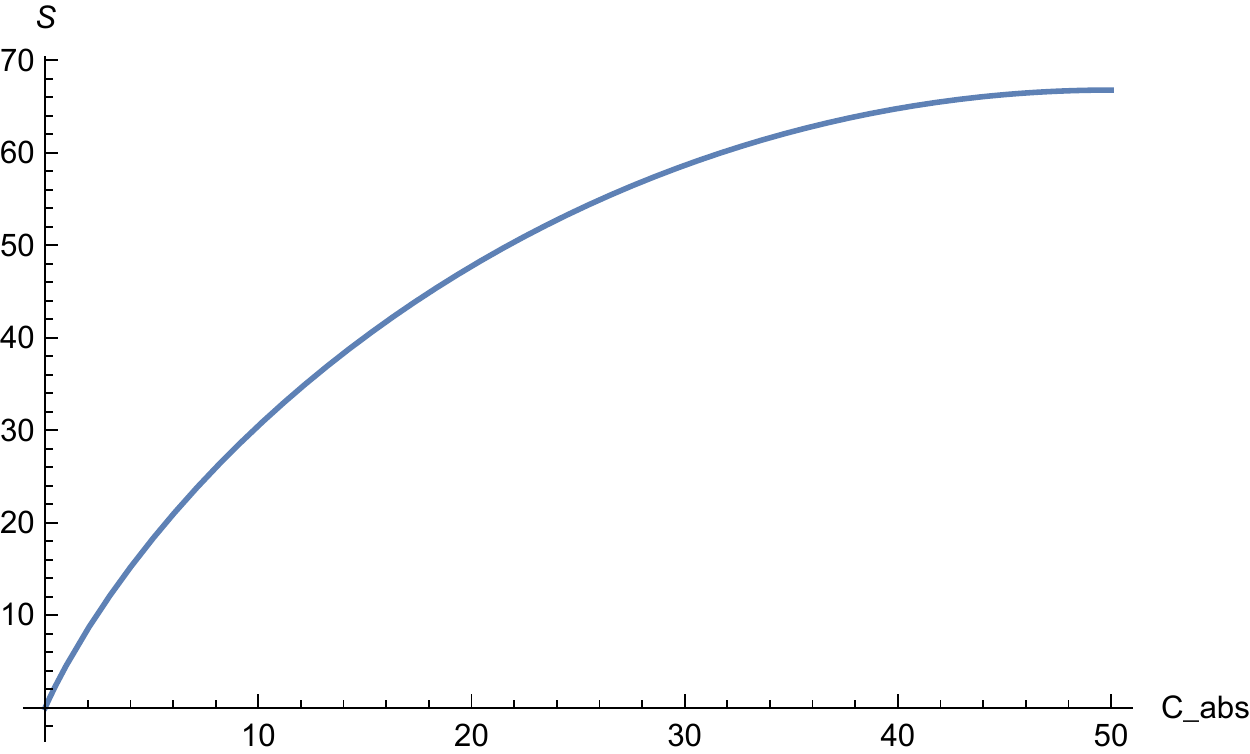}
\caption{Entropy - Absolute Complexity, $n = 2$, $K = 100$}
\label{fig:1}
\end{figure}

Figure \ref{fig:1} is a plot of eq.\eqref{eq:1}, which shows the relation between absolute complexity and entropy. Since $C_{abs} \leq \frac{K}{2}$, this function increases monotonically in the domain where $C_{abs}$ is defined. Hence, larger absolute complexity corresponds to larger entropy. At least in this case, second law of thermodynamics implies second law of classical absolute complexity: absolute complexity always tends to grow, just like entropy always tend to grow, until it reaches its maximum. We conjecture that this is true for any system where classical absolute complexity is well defined. We call this the second law of classical absolute complexity.

When our system is large, and complexity is not too small or too large, we have

\begin{equation}
\label{eq:2:1}
    S(a_0, a_1) = S(C, K) = - C \ln \frac{C}{K} - (K - C) \ln \frac{K-C}{K}
\end{equation}

If we look at the absolute complexity, then \eqref{eq:2:1} becomes

\begin{equation}
    S(C_{abs}, K)  = -2 C_{abs} \ln \frac{C_{abs}}{K}
\end{equation}

When $C << K$, 

\begin{equation}
\label{eq:1}
    S(C,K) = C \ln K - C \ln C \approx C \ln K
\end{equation}

When $C \approx C_{abs} = \frac{K}{2}$,

\begin{equation}
    S(C,K) \approx C \ln 4
\end{equation}

\section{Growth of classical complexity}
\label{sec:4}

We are interested in the growth of classical complexity, under operations. Note that we can have different protocols in applying our operators. Naturally, the larger the $n$, the more complicated the system is. We first discuss about the features when $n = 2$ (the bit model) and $n = 3$ (the trit model), and then move on to general $n$'s (the n-dit model).

\subsection{The bit model}

When $n = 2$, we operate on bits and $T_+ = T_-$. Since the identity operators are boring, our protocol is: we start at the "all-zero" state; at each time step, we pick a random bit, and act $T_+$ on it, which changes the complexity of that bit and also the system by one.

\subsubsection{General feature}

We can calculate the probability-averaged complexity growth (the expectation value of complexity over time), denoted by $ C(t) $, analytically. To distinguish from the probability-averaged complexity, the actual complexity at time $t$ (which may vary in different trials) is denoted by $C_t$. Since $C = a_1$, the probability of complexity growing by one is 

\begin{equation}
\label{eq:4:1}
    Pr_{+}(C) = 1 - \frac{a_1}{K} = 1- \frac{C}{K}
\end{equation}

The probability of complexity decreasing by one is 

\begin{equation}
\label{eq:4:2}
    Pr_{-}(C) = \frac{a_1}{K} = \frac{C}{K}
\end{equation}

$C(0) = 0$, and 
\begin{equation}
     C(t + 1)  = Pr_{+} \big( C(t) \big) \big( C(t)  + 1 \big) + Pr_{-} \big( C(t) \big) \big( C(t)  - 1\big),
\end{equation}
which can be simplified as 
\begin{equation}
\label{lya}
     C(t + 1)  = \Big(1 - \frac{2}{K}\Big)  C(t)  + 1
\end{equation}

After solving the difference equation, we get the growth of average complexity 

\begin{equation}
\label{eq:Ct}
     C(t)  = \frac{K}{2} - \frac{K}{2} \Big(1 - \frac{2}{K}\Big)^t
\end{equation}

Note that $C(t) \leq \frac{K}{2}$, and in this regime, $C = C_{abs}$, eq.\eqref{eq:Ct} is also the growth of average absolute complexity.

\begin{figure}[tbp]
\centering
\includegraphics[scale = 0.75]{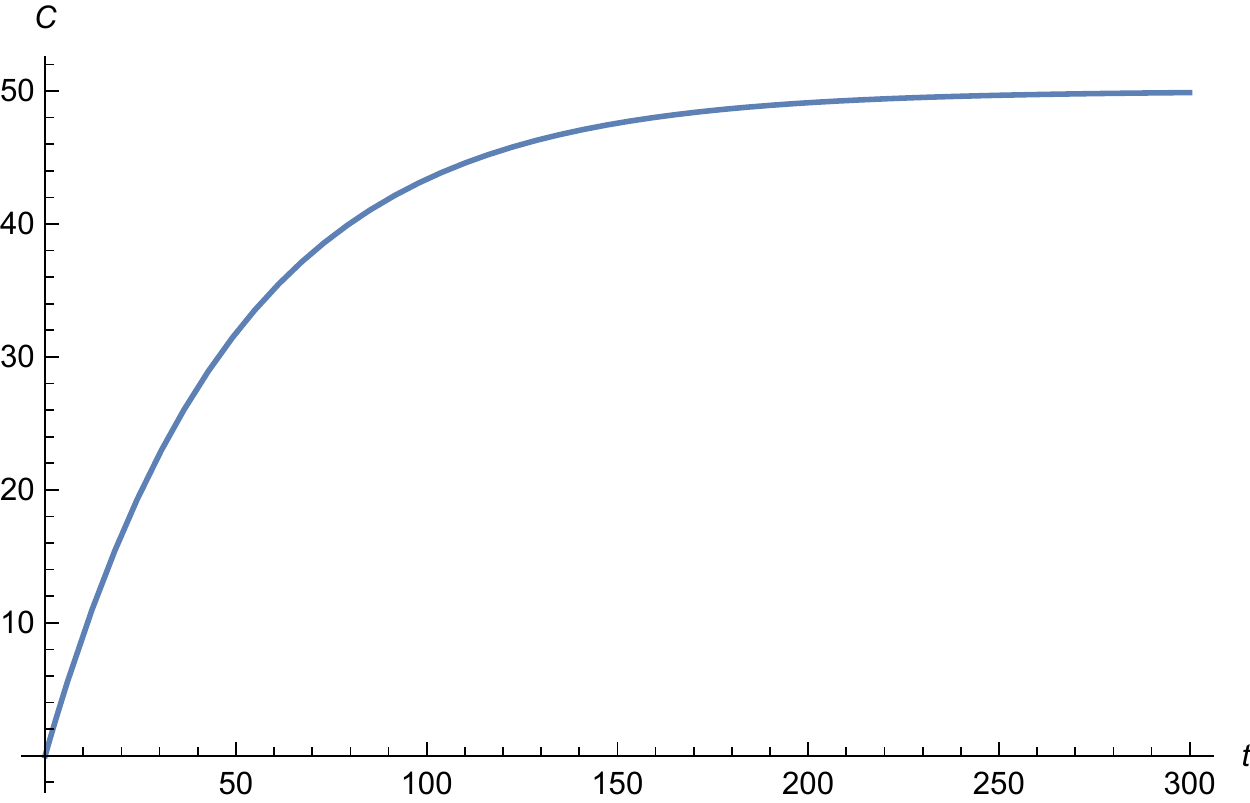}
\caption{Complexity - Time, $n = 2$, $K = 100$}
\label{fig:2}
\end{figure}

Figure \ref{fig:2} is a plot of \eqref{eq:Ct}. As $t$ becomes very large, which refers to late time, complexity fluctuates around $\frac{K}{2}$, which corresponds to the states with the maximal absolute complexity.

When $K$ is very large, it is unlikely to operate on one bit for more than once, so we would expect a linear growth of complexity.

\begin{equation}
     C(t) = t
\end{equation}

We could also see this from Eq.(4.5), since for large $K$,

\begin{equation}
     C(t) \approx \frac{K}{2} - \frac{K}{2} \Big(1 - \frac{2t}{K}\Big) = t
\end{equation}

We may wonder how long it takes to reach the maximal absolute complexity, i.e. the complexity equilibrium. A rough estimate comes from the Lyapunov exponent \cite{goldstein_classical_2002,bolotin_chaos:_2016}.

Eq.\eqref{lya} generates a series of points on the real axis, the corresponding Lyapunov exponent is

\begin{equation}
    \lambda = \lim_{t \to \infty} \frac{1}{t} \sum_{k = 0}^{t - 1} \ln \Big(1 - \frac{2}{K}\Big) = \ln \Big(1 - \frac{2}{K}\Big) < 0
\end{equation}

The negative sign indicates that these points will reach a fixed point or a superstable periodic orbit. In our case, the points will converge to $\frac{K}{2}$. The time it takes to reach maximal absolute complexity is of order 

\begin{equation}
    \tau \sim \frac{1}{|\lambda|} = \frac{-1}{\ln \Big(1 - \frac{2}{K}\Big)}
\end{equation}

When $K$ is large, $\tau \sim \frac{K}{2}$.

This is a very rough estimate. In next section, we will gain furthur understanding of the model by looking at it from the Markovian perspective.
We will show two ways of calculating the recurrence time, as well as two ways of calculating $\tau$.

\subsubsection{Markovian perspective}

There are $2^K$ different configuration states of the system, each of which is a string of "0"s and "1"s with length $K$. We can label them as $s_1$, $s_2$, $s_3$, ... , $s_{2^K}$. Since each time we act on a random bit, the configuration state of the system at time $t + 1$ is only dependent on the configuration state of the system at time $t$, so the process is Markovian. Since we have $2^K$ different configuration states, the transition matrix is $2^K \times 2^K$. Starting from an arbitrary state, after one bit-flip operation, it will end up in one of the $K$ possible states, with probability $\frac{1}{K}$. Hence, each row of the transition matrix has $K$ non-zero entries, all of which have value $\frac{1}{K}$. 
The transition matrix is ergodic, since it is possible to go from an arbitrary state to another arbitrary state, though not necessarily in one move. 
Each state is on an equal footing, so the fixed vector of the transition matrix is a row vector with $2^K$ entries, all of which have value $r = \frac{1}{2^K}$. The recurrence time of an arbitrary state is $\frac{1}{r} = 2^K$. Since we start from the "all-zero" state, which is the only state of complexity 0, the recurrence time of complexity is also $2^K$. 

While this approach is straightforward and intuitive, the transition matrix is too large to handle. Now we introduce another approach.

From equations \eqref{eq:4:1} and \eqref{eq:4:2}, we realize that the complexity of the system at the next time step is only dependent on the complexity at the moment, and we don't need other information of the system. For example, the value of the first bit doesn't matter. Hence, if we define state $i$ as the state with complexity $i$, then we have established a Markov chain. We call these specially defined states "complexity state", to distinguish from the states we usually think of, e.g. the configuration states at the beginning of this subsection.

The transition probabilities are:
\begin{equation}
    p_{ij} = 1 - \frac{i}{K}, \ j = i+1
\end{equation}
\begin{equation}
    p_{ij} = \frac{i}{K}, \ j = i-1
\end{equation}
\begin{equation}
    p_{ij} = 0, \ j \neq i \pm 1
\end{equation}

Since it is possible to reach any possible complexity starting from any complexity state, the Markov chain is ergodic. 

How are the configuration states and the complexity states related? The number of configuration states with complexity $C$ is

\begin{equation}
    N(C) = \binom{K}{C}
\end{equation}

Hence, each complexity state corresponds to different number of configuration states. For example, the "all-zero" state is the only configuration state that has complexity 0, but there are $K$ different configuration states that have complexity 1. We would expect that the proportion of time spent in complexity state $i$, denoted by $w_i$, is the ratio between the number of configuration states of complexity and the number of total states:

\begin{equation}
\label{eq:4:!4}
    w_i = \frac{N(i)}{2^K} = \frac{\binom{K}{i}}{2^K}
\end{equation}

The $w_i$'s form a row vector $w$. We now show that $w$ is actually the fixed vector of the transition matrix, also called the stationary distribution of the Markov chain. We show this by using Theorem 6 of the appendix.

\textit{Proof}. It is sufficient to show that $w_i p_{ij} = w_j p_{ji}$. This is immediate when $|i - j| \neq 1$ since both sides are zero. Now we only need to show that $w_i p_{i (i+1)} = w_{i+1} p_{(i+1)i}$.

\begin{equation}
    w_i p_{i (i+1)} - w_{i+1} p_{(i+1)i} = \frac{(K-1)!}{2^K} \Bigg(\frac{K-i}{i!(K-i)!}-\frac{i+1}{(i+1)!(K-i-1)!}\Bigg)=0.
\end{equation}

Hence $w$ is the fixed vector. The system satisfies detailed balance. The mean recurrence time for the "all-zero" state (the unique state with complexity 0) is 

\begin{equation}
    r_0 = \frac{1}{w_0} = 2^K
\end{equation}
which agrees with the result from our previous discussion on configuration states.

The complexity equilibrium is also illustrated by the stationary distribution $w$. When $i = \frac{K}{2}$, $w_i$ reaches its maximal value, which corresponds to the longest occupation time. So at late time, it is most likely to find our system in complexity state $\frac{K}{2}$. The average fluctuation of complexity at late time (the standard deviation) is

\begin{equation}
    \Delta C = \sqrt{\sum_{i=0}^{K} w_i \Big(i - \frac{K}{2}\Big)^2}= \frac{\sqrt{K}}{2}
\end{equation}

Our main goal is to calculate $\tau$. With all these preparations, now we are ready to proceed.

\textbf{Method I (The Numerical Method)}:
We write down the transition matrix $P$ based on Eq.(4.10)-(4.12). Then what we need is the mean first passage time from the complexity 0 state to the complexity $\frac{K}{2}$ state. This can be calculated numerically using Theorem 8 in the appendix, since

\begin{equation}
    \tau = m_{0\frac{K}{2}}
\end{equation}

All we need to do is to calculate $m_{0 \frac{K}{2}}$ based on $P$. The only limitation is that when $K$ grows large, it takes more time to run the program. The advantage is that this method is very general, which can be used to calculate the time it takes to go from any complexity state to another complexity state.

\textbf{Method II (The Analytical Method)}: This is done by mathematician Gunnar Blom \cite{blom_mean_1989}. Here we cite the result, 

\begin{equation}
    \tau = \frac{K}{2} \sum_{j=0}^{K/2\ -1} \frac{1}{2 j + 1}
\end{equation}

This can be approximated by an integral

\begin{equation}
    \tau \approx \frac{K}{2} \int_{0}^{K/2 \ - 1} \frac{1}{2x + 1} dx = \frac{K}{4} \ln (K - 1) 
\end{equation}

\subsection{The trit model}

When $n = 3$, we operate on trits. Unlike when we have bits, here $T_{+} \neq T_{-}$. Hence, we can have more interesting protocols. The following protocol is defined as time-independent:

At each time step, we pick a random trit, and there is probability $q_0$ that it is acted by the identity operator $I$, probability $q_1$ that it is acted by $T_+$, and probability $q_2$ that it is acted by $T_-$, where $q_0$, $q_1$ anf $q_2$ are constants, i.e., time-independent.

There are $3^K$ different configuration states of the system. Suppose our protocol is time-independent. Since at each time step, we act on a random trit, and the individual state of the trit after the operation is only dependent on the original individual state and some constants, the configuration state at time $t+1$ is only dependent on the configuration state at time $t$, so the process is Markovian. Using a similar argument to the case when $n = 2$, we conclude that the recurrence time of zero complexity is $3^K$.

Since the identity operators are boring, we mainly consider the following protocols:

1. At each time step, we pick a random trit, and act $T_+$ ($T_-$) on it.

2. At each time step, we pick a random trit, and there is $\frac{1}{2}$ probability that we act $T_+$ on it, and $\frac{1}{2}$ probability that we act $T_-$ on it.

Note that when $n = 2$, these two protocols are identical.

Now we calculate the growth of complexity for these two protocols. For an arbitrary trit, it has three possible states: 0, 1, and 2. Suppose the probability of the trit being at state $i$ is $u_i$, we assign a probability vector $u(t) = (u_0(t), u_1(t), u_2(t))$ to represent the probability distribution. The expectation value of the number of trits at state $i$ is $a_i(t) = K u_i(t)$. If we focus on the individual state of a given trit, the process is Markovian. $P_+$ ($P_-$) is the transition matrix if we only use operator $T_+$ ($T_-$).

\begin{equation}
    P_+ = \left(
\begin{array}{ccc}
 1-\frac{1}{K} & \frac{1}{K} & 0 \\
 0 & 1-\frac{1}{K} & \frac{1}{K} \\
 \frac{1}{K} & 0 & 1-\frac{1}{K} \\
\end{array}
\right)
  \   P_- = \left(
\begin{array}{ccc}
 1-\frac{1}{K} & 0 & \frac{1}{K} \\
 \frac{1}{K} & 1-\frac{1}{K} & 0 \\
 0 & \frac{1}{K} & 1-\frac{1}{K} \\
\end{array}
\right)
\end{equation}

$P_2$ is the transition matrix for protocol 2.

\begin{equation}
    P_2 = \frac{1}{2} (P_+ + P_-) = \left(
\begin{array}{ccc}
 1-\frac{1}{K} & \frac{1}{2 K} & \frac{1}{2 K} \\
 \frac{1}{2 K} & 1-\frac{1}{K} & \frac{1}{2 K} \\
 \frac{1}{2 K} & \frac{1}{2 K} & 1-\frac{1}{K} \\
\end{array}
\right)
\end{equation}

\begin{figure}[tbp]
\centering
\includegraphics[scale = 0.5]{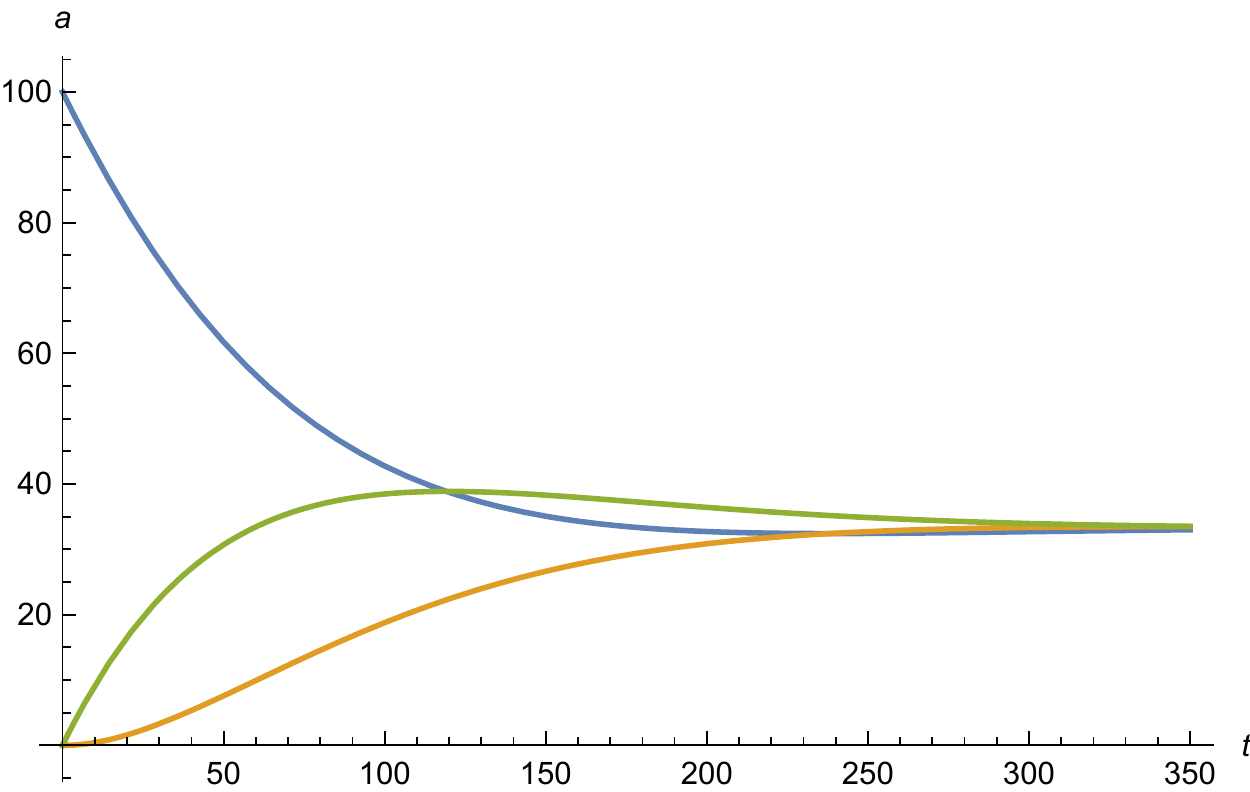}
\includegraphics[scale = 0.5]{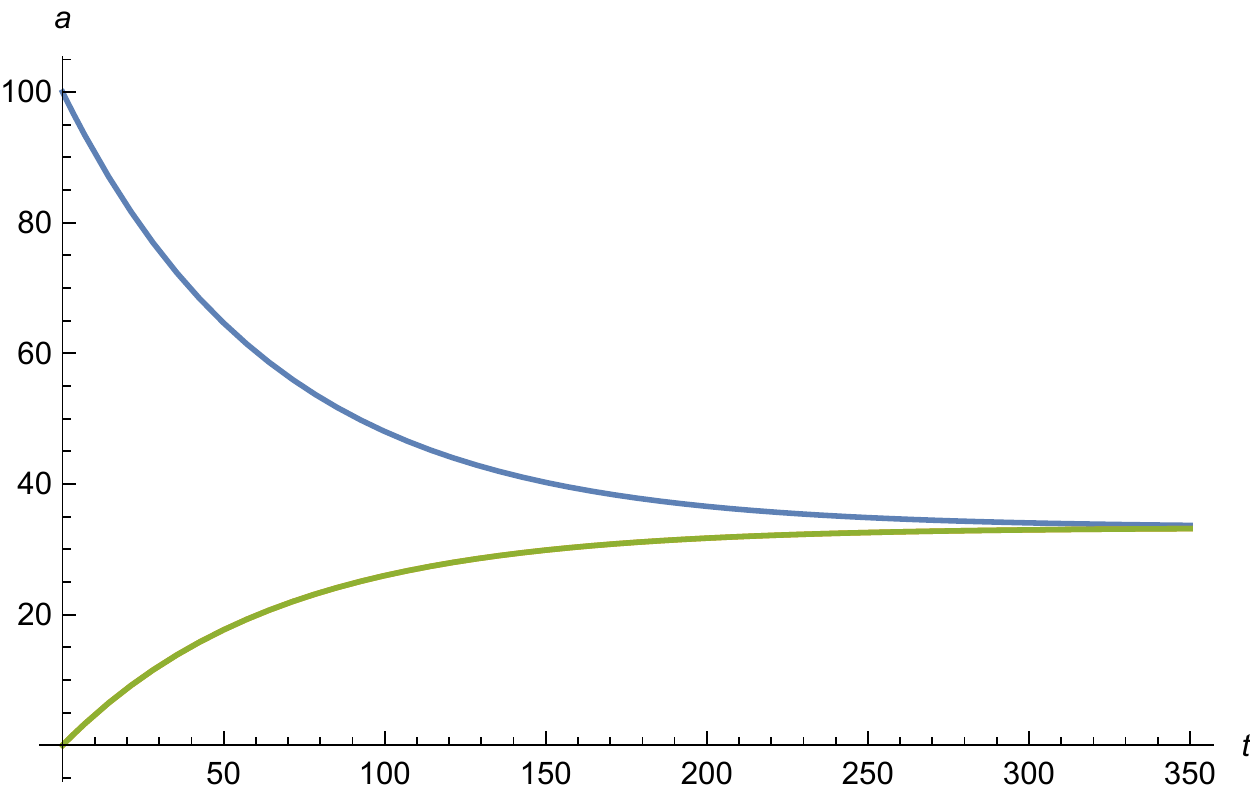}
\caption{Number of trits at state $i$ - Time, $n = 3$, $K = 100$, left graph for protocol 1, right graph for protocol 2, blue line - $a_0(t)$, green line - $a_1(t)$, orange line - $a_2(t)$}
\label{fig:3}
\end{figure}

Figure \ref{fig:3} shows the change of number of trits in each state. Regardless of protocols we use, after reaching the equilibrium, we have equal number of trits in each state. This is expected, since this is the distribution with maximal entropy and maximal absolute complexity, as discussed in section \ref{sec:3}.

Since $u(t) = u(0) P^t$, $u(0) = (1, 0, 0)$, and $C(t) = a_1(t) + a_2(t) = K u_1(t) + K u_2(t)$, we calculate the classical complexity as a function of time.

\begin{equation}
\label{eq:3:1}
   C_1(t) = C_+(t) = C_-(t) = \frac{2K}{3} - \frac{K}{3} \Bigg( \bigg(1 - \frac{3}{2K} + \frac{\sqrt{3}i}{2K} \bigg)^t + \bigg(1 - \frac{3}{2K} - \frac{\sqrt{3}i}{2K} \bigg)^t \Bigg)
\end{equation}

\begin{equation}
\label{eq:3:2}
    C_2(t) = \frac{2K}{3} - \frac{2K}{3} \Big(1 - \frac{3}{2K}\Big)^t
\end{equation}

\begin{figure}[tbp]
\centering
\includegraphics[scale = 0.75]{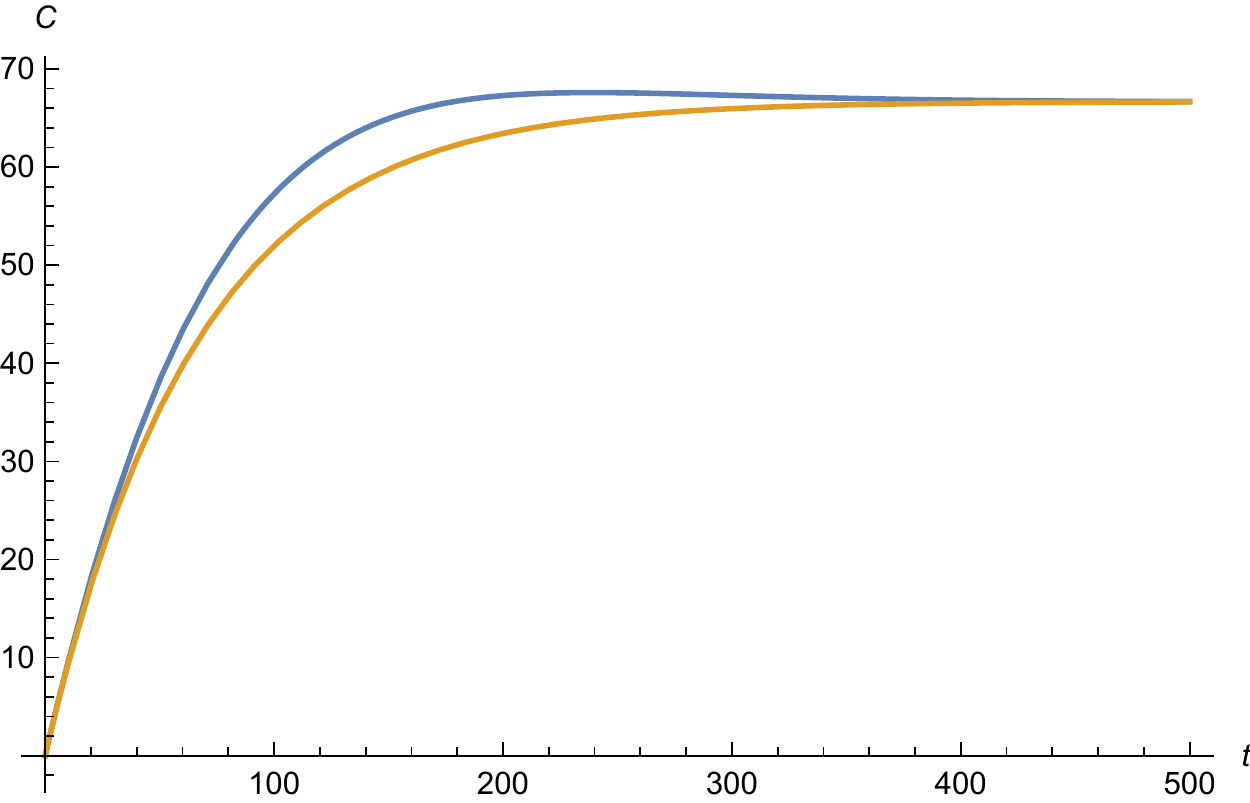}
\caption{Complexity - Time, $n = 3$, $K = 100$, blue line for protocol 1, orange line for protocol 2}
\label{fig:4}
\end{figure}

Figure \ref{fig:4} is a plot of equations \eqref{eq:3:1} and \eqref{eq:3:2}. Note that under protocol 1, complexity first exceeds maximal absolute complexity and then reaches equilibrium. This "overshoot" behavior is the key difference between complexity and absolute complexity, as average absolute complexity only monotonically increases over time.

\subsection{The n-dit model}

When we have an arbitrary $n$, there are $n^K$ different configuration states of the system. When our protocol is time-independent, by the exactly same reasoning as above, the recurrence time of zero  complexity is $n^K$.

The two interesting protocols discussed above also apply here. The entries of $P_+$ are $p_{+ii} = 1 - \frac{1}{K}$ for any $i$, $p_{+i(i+1)} = \frac{1}{K}$ for $i \neq n$, $p_{+n1} = \frac{1}{K}$, and the rest are all zero.

\begin{figure}[tbp]
\centering
\includegraphics[scale = 0.5]{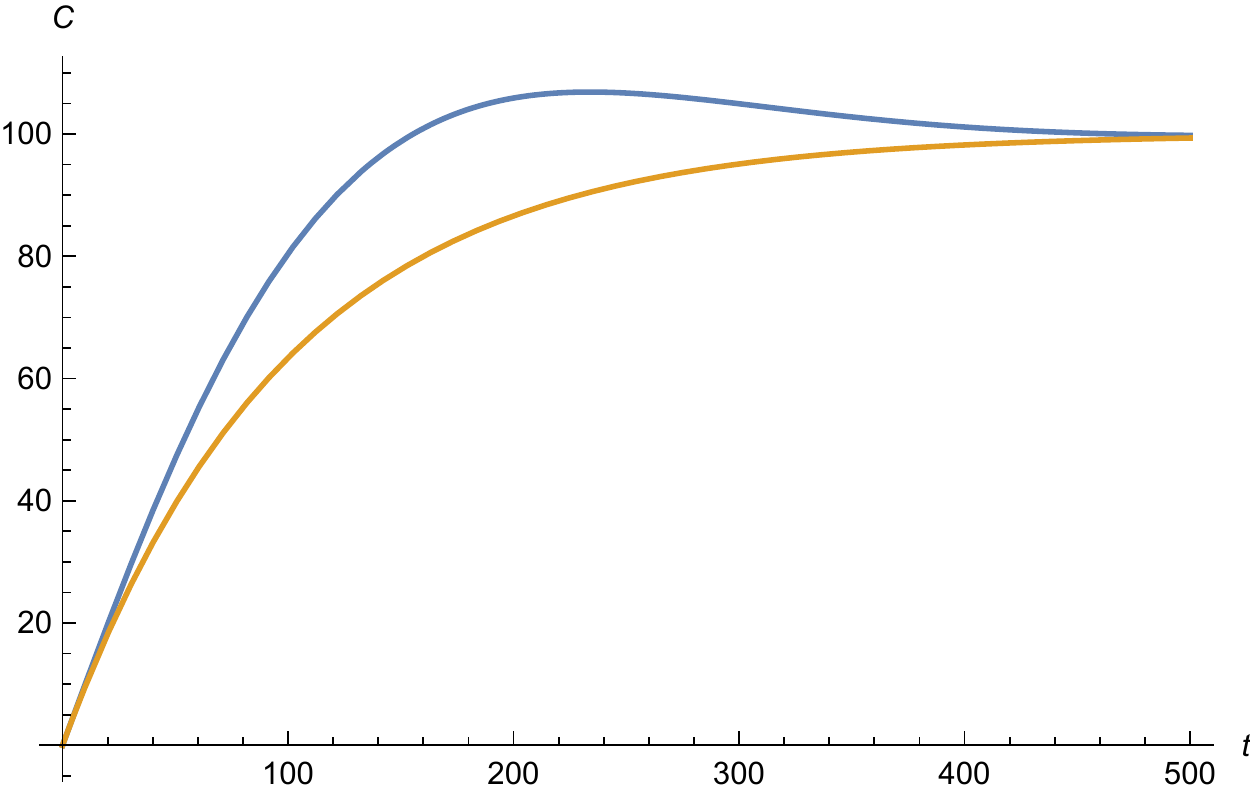}
\includegraphics[scale = 0.5]{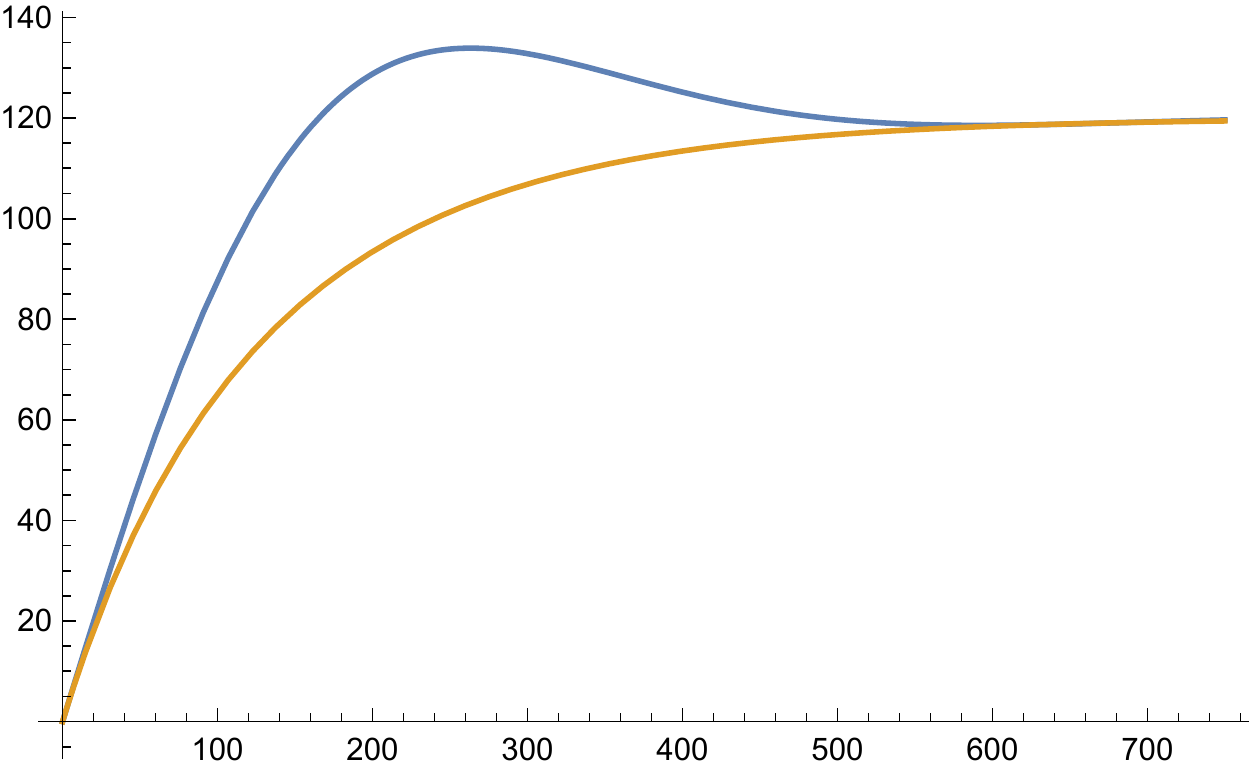}
\caption{Complexity - Time, left graph for $n = 4$, right graph for $n = 5$, $K = 100$, blue line for protocol 1, orange line for protocol 2}
\label{fig:extra}
\end{figure}

Using the same methods as discussed above, we can calculate and plot the growth of complexity for any $n$. Figure \ref{fig:extra} shows the case when $n = 4$ and $n = 5$, and we could see that the "overshoot" behavior becomes more significant as $n$ becomes larger. 

For protocol 2, due to symmetry, the growth of complexity $C(t)$ at each time step is only dependent on the number of n-dits with the minimal and maximal individual complexities.

When $n$ is even,
\begin{equation}
    C(t+1) - C(t) = \frac{1}{K} \Big(a_0(t) - a_{n/2}(t)\Big)
\end{equation}

When $n$ is odd,
\begin{equation}
    C(t+1) - C(t) = \frac{1}{K} \Big(a_0(t) - \frac{1}{2} a_{(n-1)/2}(t) - \frac{1}{2} a_{(n+1)/2}(t)\Big)
\end{equation}

At early time, it is unlikely to have a n-dit with the largest possible individual complexity, since it requires multiple operations on a individual n-dit out of the total $K$ n-dits. Hence, it is a very good approximation to have

\begin{equation}
    C(t+1) - C(t) = \frac{a_0(t)}{K}
\end{equation}

When $K$ is large, at early time, mostly likely we will act on different n-dit at each time step, so 

\begin{equation}
    b_0(t) = K - t
\end{equation}

Since $C(0) = 0$, 
\begin{equation}
    C(t) = - \frac{1}{2K} t^2 + \Big(1-\frac{1}{2K}\Big)t
\end{equation}

\begin{figure}[tbp]
\centering
\includegraphics[scale = 0.75]{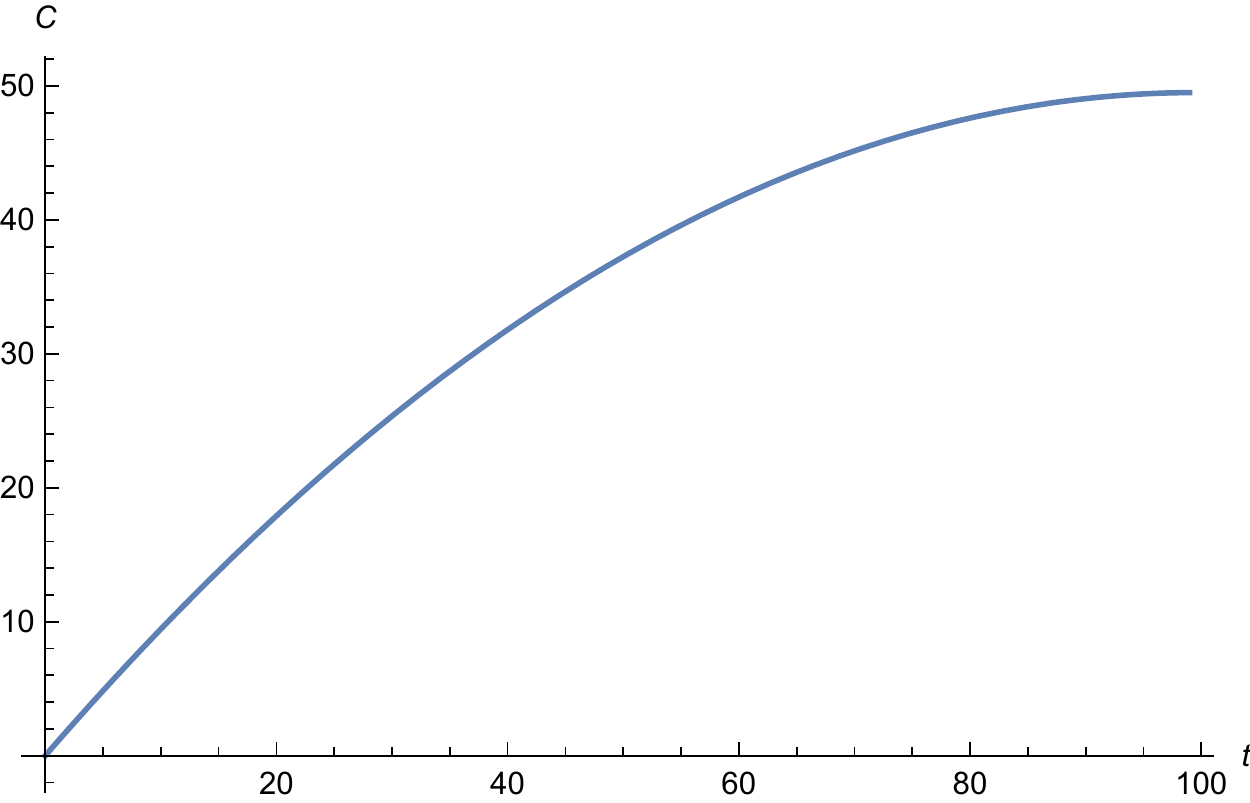}
\caption{Complexity - Time, general $n$, $K = 100$}
\label{fig:5}
\end{figure}

This approximation breaks down after $C(t)$ reaches its maximum. The time it takes to reach the maximum is $K - \frac{1}{2}$, which is a good approximation of $\tau$, the time it takes to reach the maximal absolute complexity.

\section{Growth of quantum complexity}
\label{sec:5}

As long as the probability of complexity growth is only dependent on the current complexity, the time evolution of complexity forms a Markov chain. It doesn't matter if we are considering classical or quantum complexity. 

For a simple random circuit model with only one gate, at each time step, there are $\binom{K}{2} = \frac{K(K-1)}{2}$ possible qubit pairs to act on. If we have a universal gate set, which includes $m$ gate-types that can act on any pair of qubits, then each gate involves a choice of $d = \frac{mK(K-1)}{2}$ possibilities. Brown and Susskind \cite{brown_second_2018} state that collisions are rare until late time. 

Hence, in early time, at each time step, the probability of complexity growing by one is 

\begin{equation}
    Pr_+ (C) = 1 - \frac{1}{d}
\end{equation}

The probability of complexity decreasing by one is 

\begin{equation}
    Pr_- (C) = \frac{1}{d}
\end{equation}

Note that here the probabilities are constant functions of current complexity. 

$C(0) = 0$, and

\begin{equation}
     C(t + 1) - C(t) = Pr_{+} \big( C(t) \big) - Pr_{-} \big( C(t) \big) = 1 - \frac{2}{d}
\end{equation}

This difference equation is easy to solve, and we get the growth of expectation value of complexity

\begin{equation}
    C(t) = \Big(1 - \frac{2}{d}\Big)t
\end{equation}

The average time it takes to reach complexity $C$ is 

\begin{equation}
    \tau(C) = \frac{d C}{d-2} \approx \Big(1 + \frac{2}{d} \Big) C
\end{equation}

We can see that $C(t)$ grows linearly with $t$, and when $K$ is large, $d$ is large, $C(t) = t$. 

How about the growth behavior at late time when collisions occur? A simple toy model we can consider is that collisions are very rate until the complexity reaches $C_{max} - h$. Then this late time region looks very similar to our classical complexity model, which glues the early time linear growth region and the equilibrium region smoothly together.

If we are only interested in the first passage time to maximal quantum complexity $C_{max}$, then it is the sum of two parts, the first passage time $\tau_e$ from complexity state 0 to complexity state $C_{max} - h$, and the first passage time $\tau_l$ from complexity state $C_{max} - h$ to complexity state $C_{max}$. 

\begin{equation}
    \tau_e = \frac{d (C_{max} - h)}{d-2}
\end{equation}

$\tau_l$ can be calculated from our methods discussed in Section \ref{sec:4}.

A toy example is when $h = 1$. From this assumption, we can see that it is very unlikely to collide with a state with complexity less than $C_{max} - 2$. Suppose that at the next time step, the probability of complexity increasing by one is $p_+$, the probability of complexity decreasing by one is $p_-$, the probability of complexity staying the same is $p_0$, and $p_+ + p_0 + p_- = 1$, then using conditional probability,

\begin{equation}
    \tau_l = p_+ + p_0 \tau_l + p_- (\tau_l + 1)
\end{equation}

Solve the equation and we get 

\begin{equation}
    \tau_l = 1 + \frac{p_-}{p_+}
\end{equation}

A comparison of $\frac{2}{d}$ and $\frac{p_-}{p_+}$ can help indicate the proper value of $h$. The values of $p_+$, $p_-$ and $p_0$, are dependent on the available states.

\section{Classical complexity equals non-interacting energy: the case of Ising model}
\label{sec:6}

So far we have discussed classical complexity in idealized systems with dits, we may wonder how to realize these systems. A good candidate would be the Ising model. 

Suppose we have a $d$-dimensional lattice with $K$ sites. At each site $k$ of the lattice, there is a spin $\sigma_k$ which takes the two values of $+1$ or $-1$. A spin configuration is an assignment of spin value to each lattice site. The energy of a configuration $\sigma$ is

\begin{equation}
    H(\sigma) = -J \sum_{\langle i j \rangle} \sigma_i \sigma_j - h \sum_j \sigma_j
\end{equation}

Here, $\langle i j \rangle$ indicates that sites $i$ and sites $j$ are nearest neighbors, $J$ indicates the correlation strength, and $h$ indicates the external magnetic field.

If the given simplest state is the state with all spins taking value $-1$, then the second term of $H(\sigma)$ can be directly related to the classical complexity of the system:

\begin{equation}
   -h \sum_{j} \sigma_j = -h (C + (-1) (K-C)) = - 2 h C + h K
\end{equation}

This non-interacting energy term is essentially a term of classical complexity. If we turn off the correlation, which means that we take $J = 0$, then the partition function is 

\begin{equation}
    Z = \exp(-\beta h K) \sum_{C = 0} ^ { K} \binom{K}{C} \exp(2 \beta h C)
\end{equation}

The probability of being a state with complexity $C$ is 

\begin{equation}
    P(C) = \frac{\binom{K}{C} \exp(2 \beta h C)}{\sum_{C = 0} ^ {K} \binom{K}{C} \exp(2 \beta h C)}
\end{equation}


Note that when $\beta = 0$, this simplifies to

\begin{equation}
    P(C) = \frac{\binom{K}{C}}{2^K}
\end{equation}
which is in agreement with eq. \eqref{eq:4:!4}. 

The first term of $H(\sigma)$ is also related to classical complexity, though in a less obvious way. For each pair $(i, j)$ of nearest neighbor, $\sigma_i \sigma_j = 1$ if they have the same individual complexity, and $\sigma_i \sigma_j = -1$ if they don't have the same individual complexity. If we calculate the absolute complexity of the pair, then 

\begin{equation}
    C_{abs} (\sigma_i \sigma_j = 1) = 0 = \min(C_{abs})
\end{equation}

\begin{equation}
    C_{abs} (\sigma_i \sigma_j = -1) = 1 = \max(C_{abs})
\end{equation}

This shows that the spin-spin correlation term acts against the growth of classical absolute complexity. When $\beta = 0$, this term doesn't play a role in calculating $P(C)$.

Hence, the model of bits is dual to the Ising model at infinite temperature. The example of Ising model shows that at finite temperature, classical (absolute) complexity doesn't always grow. This is not too surprising. Just like there is a battle between energy minimization and entropy maximization, similar things happen when complexity is involved, which hint towards a possible first law of classical complexity.

\acknowledgments

I would like to thank Adam Brown, Le Hu, Edward Mazenc, Brandon Rayhaun, Zhengyan Shi and Sheng Zhong for helpful comments. I also thank Leonard Susskind for numerous discussions and encouragements.

\appendix
\section{Markov chain}
We describe a Markov chain as follows: Suppose we have a set of states, $S = \{ s_1,s_2,...,s_r \}$. 
The process starts in one of these states and moves successively from one state to another. Each move is called a step. 
If the chain is currently in state $s_i$, then the probability of it moving to state $s_j$ at the next step is $p_{ij}$, and $p_{ij}$ does not depend on which states the chain was in before the current state.

The probabilities $p_{ij}$ are called transition probabilities. A initial probability distribution, defined on $S$, specifies the starting state. 

\subsection{Transition Matrix}

The matrix whose $ij$-th entry is the transition probability $p_{ij}$ is called the transition matrix. or the matrix of transition probabilities.

\begin{flushleft}
\textbf{Theorem 1} Let $P$ be the transition matrix of a Markov chain. The $ij$-th entry $p_{ij}^{(n)}$ of the matrix $P^n$ gives the probability that the Markov chain, starting in state $s_i$, will be in state $s_j$ after n steps.
\end{flushleft}

\begin{flushleft}
\textbf{Theorem 2} Let $P$ be the transition matrix of a Markov chain, and let $u$ be the probability vector which represents the starting distribution. Then the probability that the chain is in state $s_i$ after $n$ steps is the $i$-th entry in the vector
\begin{equation}
    u^{(n)} = u P^n
\end{equation}
\end{flushleft}

\subsection{Ergodic Markov Chain}

\begin{flushleft}
\textbf{Definition 1} A Markov chain is called an ergodic chain if it is possible to go from every state to every state (not necessarily in one move).
\end{flushleft}

Ergodic Markov chains are also called irreducible.

\begin{flushleft}
\textbf{Definition 2} A Markov chain is called a regular chain if some power of the transition matrix has only positive elements.
\end{flushleft}

\begin{flushleft}
\textbf{Definition 3} A row vector $w$ with the property $wP = w$ is called a fixed row vector for $P$.
\end{flushleft}

$w$ is called the stationary distribution of the Markov chain. The following theorems illustrate its properties.

\begin{flushleft}
\textbf{Theorem 3} For an ergodic Markov chain, there is a unique probability vector $w$ such that $wP = w$ and $w$ is strictly positive. Any row vector such that $vP = v$ is a multiple of $w$. 
\end{flushleft}

\begin{flushleft}
\textbf{Theorem 4} Let $P$ be the transition matrix for an ergodic chain. Let $A_n$ be the matrix defined by 
\begin{equation}
    A_n = \frac{I + P + P^2 + \dots + P^n}{n + 1}
\end{equation}
Then $A_n \to W$, where $W$ is a matrix all of whose rows are equal to the unique fixed probability vector $w$ for $P$.
\end{flushleft}

\begin{flushleft}
\textbf{Theorem 5 (Law of Large Numbers for Ergodic Markov Chains)} Let $H_j^{(n)}$ be the proportion of times in $n$ steps that an ergodic chain is in state $s_j$. Then for any $\epsilon > 0$,

\begin{equation}
    P \Big(|H_j^{(n)} - w_j| > \epsilon \Big) \to 0
\end{equation}
independent of the starting state $s_i$.
\end{flushleft}

\subsection{Detailed Balance}

\begin{flushleft}
\textbf{Definition 4}
A Markov chain with fixed vector $w$ is said to be reversible or to satisfy detailed balance (with respect to $w$) if
\begin{equation}
\label{A:4}
    w_i p_{ij} = w_j p_{ji} 
\end{equation}
for all $i$, $j$.

\end{flushleft}
Eq. \eqref{A:4} are called the detailed balance equations. Note that this is stronger than the condition that $w$ is a fixed vector (stationary distribution). Sometimes this latter system is called the "global balance equations". 

\begin{flushleft}
\textbf{Theorem 6} Let $P$ be the transition matrix for a Markov chain, and suppose there exists a row vector $w$ such that $ w_i p_{ij} = w_j p_{ji} $ for all allowed $i$, $j$. Then $w$ is a fixed vector of the chain, and the chain is reversible.

\textit{Proof}. Suppose $w$ satisfies the conditions of the theorem. Then

\begin{equation}
    \sum_{i} w_i p_{ij} = \sum_{i} w_j p_{ji} = w_j \sum_i p_{ji} = w_j
\end{equation}

Hence, $w = wP$, $w$ is a fixed vector. From Definition 4, we know that the chain is reversible.

\end{flushleft}

\subsection{Mean Recurrence Time}

\begin{flushleft}
\textbf{Definition 5} If an ergodic Markov chain is started in state $s_i$, the expected number of steps to return to $s_i$ for the first time is the mean recurrence time for $s_i$. It is denoted by $r_i$.
\end{flushleft}

\begin{flushleft}
\textbf{Theorem 7} For an ergodic Markov chain, the mean recurrence time for state $s_i$ is $r_i = \frac{1}{w_i}$, where $w_i$ is the $i$-th component of the fixed probability vector for the transition matrix.
\end{flushleft}

\subsection{Mean First Passage Time}

\begin{flushleft}
\textbf{Definition 6} If an ergodic chain is started in the state $s_i$, the expected number of steps to reach state $s_j$ for the first time is called the mean first passage time from $s_i$ to $s_j$. It is denoted by $m_{ij}$. By convention $m_{ii}$ = 0.
\end{flushleft}

\begin{flushleft}
\textbf{Definition 7 (Fundamental Matrix)} Let $P$ be the transition matrix of an ergodic chain, and let $W$ be the matrix all of whose rows are the fixed probability row vector for $P$. Then the matrix 
\begin{equation}
    Z = (I - P + W)^{-1}
\end{equation}
is the fundamental matrix of the ergodic chain.
\end{flushleft}

\begin{flushleft}
\textbf{Definition 8 (Mean First Passage Matrix)} The mean first passage matrix, denoted by $M$, is a matrix whose $ij$-th entry $m_{ij}$ is the mean first passage time to go  from $s_i$ to $s_j$ if $i \neq j$; the diagonal entries are 0.
\end{flushleft}

\begin{flushleft}
\textbf{Theorem 8} The mean first passage time matrix $M$ for an ergodic chain is determined from the fundamental matrix $Z$ and the fixed row probability $w$ by
\begin{equation}
    m_{ij} = \frac{z_{jj}-z_{ij}}{w_j}
\end{equation}
\end{flushleft}



\bibliographystyle{JHEP}
\bibliography{ref.bib}

\providecommand{\href}[2]{#2}\begingroup\raggedright\begin{thebibliography}{10}

\bibitem{brown_second_2018}
A.~R. Brown and L.~Susskind, {\it The {Second} {Law} of {Quantum}
  {Complexity}},  {\em Physical Review D} {\bf 97} (Apr., 2018). arXiv:
  1701.01107.

\bibitem{susskind_switchbacks_2014}
L.~Susskind and Y.~Zhao, {\it Switchbacks and the {Bridge} to {Nowhere}},  {\em
  arXiv:1408.2823 [hep-th, physics:quant-ph]} (Aug., 2014). arXiv: 1408.2823.

\bibitem{brown_quantum_2017}
A.~R. Brown, L.~Susskind, and Y.~Zhao, {\it Quantum {Complexity} and {Negative}
  {Curvature}},  {\em Physical Review D} {\bf 95} (Feb., 2017). arXiv:
  1608.02612.

\bibitem{lin_cayley_2019}
H.~W. Lin, {\it Cayley graphs and complexity geometry},  {\em Journal of High
  Energy Physics} {\bf 2019} (Feb., 2019).

\bibitem{susskind_three_2018}
L.~Susskind, {\it Three {Lectures} on {Complexity} and {Black} {Holes}},  {\em
  arXiv:1810.11563 [hep-th]} (Oct., 2018). arXiv: 1810.11563.

\bibitem{brown_complexity_2016}
A.~R. Brown, D.~A. Roberts, L.~Susskind, B.~Swingle, and Y.~Zhao, {\it
  Complexity, action, and black holes},  {\em Physical Review D} {\bf 93}
  (Apr., 2016). arXiv: 1512.04993.

\bibitem{brown_complexity_2016-1}
A.~R. Brown, D.~A. Roberts, L.~Susskind, B.~Swingle, and Y.~Zhao, {\it
  Complexity {Equals} {Action}},  {\em Physical Review Letters} {\bf 116} (May,
  2016). arXiv: 1509.07876.

\bibitem{susskind_computational_2014}
L.~Susskind, {\it Computational {Complexity} and {Black} {Hole} {Horizons}},
  {\em arXiv:1402.5674 [gr-qc, physics:hep-th, physics:quant-ph]} (Feb., 2014).
  arXiv: 1402.5674.

\bibitem{stanford_complexity_2014}
D.~Stanford and L.~Susskind, {\it Complexity and {Shock} {Wave} {Geometries}},
  {\em Physical Review D} {\bf 90} (Dec., 2014). arXiv: 1406.2678.

\bibitem{brown_case_2018}
A.~R. Brown, H.~Gharibyan, H.~W. Lin, L.~Susskind, L.~Thorlacius, and Y.~Zhao,
  {\it The {Case} of the {Missing} {Gates}: {Complexity} of
  {Jackiw}-{Teitelboim} {Gravity}},  {\em arXiv:1810.08741 [gr-qc,
  physics:hep-th]} (Oct., 2018). arXiv: 1810.08741.

\bibitem{goto_holographic_2018}
K.~Goto, H.~Marrochio, R.~C. Myers, L.~Queimada, and B.~Yoshida, {\it
  Holographic {Complexity} {Equals} {Which} {Action}?},  {\em arXiv:1901.00014
  [gr-qc, physics:hep-th, physics:quant-ph]} (Dec., 2018). arXiv: 1901.00014.

\bibitem{grinstead_grinstead_2009}
C.~M. Grinstead and J.~L. Snell, {\em Grinstead and {Snell}'s {Introduction} to
  probability}.
\newblock Editore, S.l., 2009.
\newblock OCLC: 898732475.

\bibitem{walsh_knowing_2012}
J.~B. Walsh, {\em Knowing the odds: an introduction to probability}.
\newblock No.~v. 139 in Graduate studies in mathematics. American Mathematical
  Society, Providence, R.I, 2012.

\bibitem{kardar_statistical_2007}
M.~Kardar, {\em Statistical physics of fields}.
\newblock Cambridge University Press, Cambridge ; New York, 2007.
\newblock OCLC: ocn123113789.

\bibitem{goldstein_classical_2002}
H.~Goldstein, C.~P. Poole, and J.~L. Safko, {\em Classical mechanics}.
\newblock Addison Wesley, San Francisco, NJ, 3. ed~ed., 2002.
\newblock OCLC: 248389949.

\bibitem{bolotin_chaos:_2016}
Y.~Bolotin, A.~Tur, and V.~Yanovsky, {\em Chaos: concepts, control and
  constructive use}.
\newblock Springer Berlin Heidelberg, New York, NY, 2016.

\bibitem{blom_mean_1989}
G.~Blom, {\it Mean transition times for the {Ehrenfest} urn model},  {\em
  Advances in Applied Probability} {\bf 21} (June, 1989) 479--480.

\end{thebibliography}\endgroup

\end{document}